\documentclass[pre,byrevtex,amsmath,amssymb,preprint]{revtex4}

\usepackage{graphicx}
\usepackage{dcolumn}
\usepackage{bm}

\begin{document}
\preprint{PREPRINT}

\title[Short Title]{Polyelectrolyte Multilayering on a Charged Sphere}

\author{Ren\'e Messina}
\email{messina@thphy.uni-duesseldorf.de}
\altaffiliation[Permanent address: ]
{Institut f\"ur theoretische Physik II,
Heinrich-Heine-Universit\"at D\"usseldorf,
Universit\"atsstrasse 1,
D-40225 D\"usseldorf,
Germany}
\affiliation{Max-Planck-Institut f\"ur Polymerforschung, Ackermannweg 10,
  55128 Mainz, Germany}
\author{Christian Holm}
\email{holm@mpip-mainz.mpg.de}
\affiliation{Max-Planck-Institut f\"ur Polymerforschung, Ackermannweg 10,
  55128 Mainz, Germany}
\author{Kurt Kremer}
\email{kremer@mpip-mainz.mpg.de}
\affiliation{Max-Planck-Institut f\"ur Polymerforschung, Ackermannweg 10,
  55128 Mainz, Germany}

\date{\today}

\begin{abstract}
  The adsorption of highly \textit{oppositely} charged flexible
  polyelectrolytes on a charged sphere is investigated by means of Monte Carlo
  simulations in a fashion which resembles the layer-by-layer deposition
  technique introduced by Decher. Electroneutrality is insured at each step by
  the presence of monovalent counterions (anions and cations).  We study in
  detail the structure of the \textit{equilibrium} complex.
  Our investigations of the first few layer formations strongly suggest that
  multilayering on a charged colloidal sphere is not possible as an
  equilibrium process with purely electrostatic interactions. We especially
  focus on the influence of specific (non-electrostatic) short range
  attractive interactions (e.g., Van der Waals) on the stability of the
  multilayers.
\end{abstract}

\maketitle

\section{Introduction}

Polyelectrolyte multilayer thin films are made of alternating layers
of polycations (PCs) and polyanions (PAs).  The so-called layer by
layer method, first introduced in \textit{planar} geometry by Decher,
consists in a successive adsorption of the polyions onto a charged
surface and has proved to be extremely
efficient\cite{Decher_1992,Decher_1997}.  Due to the many potential
technological applications such as biosensing
\cite{Caruso_Langmuir_1997}, catalysis \cite{Onda_1999}, optical
devices \cite{Wu_JACS_1999} etc., this process is nowadays widely
used.  Various techniques are employed to control the polymer
multilayer buildup such as optical \cite{Dijt_1994,Kovacevic_2002} and
neutron \cite{Schmitt_1993,Loesche_Macromol_1998} reflectometry, AFM
\cite{McAloney_2001}, UV spectroscopy \cite{Schoeler_2002}, NMR techniques
\cite{mccormick02a_pre}, and others.  Some
experiments (see e.g., Ref. \cite{Helm_Macromolecules_1998}) were devoted
to the basic mechanisms governing polyelectrolyte multilayering on
planar mica-surfaces where especially, the role of surface charge
overcompensation was pointed out.

Another very interesting application is provided by the polyelectrolyte
coating of \textit{spherical} metallic nanoparticles
\cite{Caruso_Science_1998,Caruso_JPCB_2001}.  This process can modify in a
well controlled way the physico-chemical surface properties of the colloidal
particle.  Despite of the huge amount of experimental works, the detailed
understanding of the multilayering process is still rather unclear, especially
for a charged colloidal sphere.  Hence the study of polyelectrolyte
multilayering is motivated by both experimental and theoretical interests.


On the theoretical side, the literature on this subject is rather
poor.
Based on Debye-H\"uckel approximations for the electrostatic interactions
and including lateral correlations by considering different
typical semiflexible polyelectrolyte-layer structures,
Netz and Joanny\cite{Netz_Macromol_1999b} found a remarkable
stability of the periodic structure of the multilayers in planar geometry.
%
For weakly charged flexible polyelectrolytes at high ionic
strength qualitative agreements between theory
\cite{Castelnovo_2000}, based on scaling laws, and experimental
observations \cite{Loesche_Macromol_1998,Fahrat_1999} have been
provided.
The driving force of all these multilayering processes is of
electrostatic origin and it is based on an overcharging mechanism, where
the first layer overcharges the macroion and the subsequent layers
overcharge the layers underneath.  However, the role of
non-electrostatic interactions though pointed out in
Ref. \cite{Castelnovo_2000,Joanny_EPJB_1999} is not clear.
In particular, it is still open whether the layer build up is an equilibrium-
or out of equilibrium process, which resembles more a succession of
dynamically trapped states.  Therefore we do not know whether or not the
complex polyelectrolyte is in \textit{thermodynamical equilibrium}.  This
point has also been emphasized in a recent experimental work on planar
multilayers \cite{Kovacevic_2002} where considerable kinetic effects were
reported.
%
%
So far, there are nor analytical results neither simulation data for
multilayering formation onto charged spheres.

The goal of this paper is to study the underlying physics involved in the
polyelectrolyte multilayering onto a charged colloidal sphere by means of MC
simulations.  The paper is organized as follows: Sec.  \ref{ Sec.simu_method}
is devoted to the description of our MC simulation method.  The relevant
target quantities are specified in Sec. \ref{Sec.Target}.  The single chain
adsorption is studied in Sec. \ref{Sec.single-chain}, and the polyelectrolyte
bilayering in Sec. \ref{Sec.two-chains}.  Then the multilayering process is
investigated in Sec. \ref{Sec.Multilayer}.  The case of short polyelectrolyte
chains is considered in Sec.  \ref{Sec.short-chains}. Finally, Sec.
\ref{Sec.conclu} contains some brief concluding remarks.

\section{Simulation method
 \label{ Sec.simu_method}}

The setup of the system under consideration is very similar to those recently
investigated by means of molecular dynamics simulations
\cite{Messina_PRE_2002,Messina_JCP_2002}.  Within the framework of the
primitive model we consider one charged colloidal sphere characterized by a
radius $a$ ($=4.5\sigma$) and a bare charge $Q_{M}=-Z_{M}e$ (where $e$ is the
elementary charge and $Z_{M}=40$) surrounded by $Z_M$ neutralizing monovalent 
($Z_c=1$) counterions and an implicit solvent (water)
of relative dielectric permittivity $\epsilon_{r}\approx 80$. 
In the remaining of the paper, we will refer to the term \textit{macroion} as
the charged colloidal sphere.
Polyelectrolyte chains ($N_+$ PCs and $N_-$ PAs) are made up of $N_m$
\textit{monovalent} monomers ($Z_{m}=1$) of diameter $\sigma$. For the sake of
simplicity, we only consider here symmetrical complexes where PC and PA chains
have the same length and carry the same charge in absolute value.  
To each charged PC or PA we also add $N_m$ small \textit{monovalent} ($Z_c=1$) 
counterions (anions and cations countering the charge of the polyelectrolytes) 
of diameter $\sigma$, hence always a charge neutral entity
was added. Thereby all the microions have the same valence
$Z=Z_{c}=Z_{m}=1$ as well as the same diameter $\sigma$.
Added salt of course would even weaken the effects observed and would be
especially important for the case of an adsorption interaction between
macroion and polyelectrolyte.

All these particles making up the system are confined in an impermeable
spherical cell of radius $ R=60\sigma$.  The spherical macroion is held fixed
and located at the center of the cell.  To avoid the appearance of image
charges \cite{Messina_image_2002}, we assume that the macroion has the same
dielectric constant as the solvent.

Standard canonical MC simulations following the Metropolis scheme
were used \cite{Metropolis_JCP_1953,Allen_book_1987}.
%
%
The total energy of interaction of the system can be written as

\begin{equation}
\label{eq.U_tot}
U_{tot}=\sum _{i,i<j}U_{hs}+U_{coul}+U_{FENE}+U_{LJ}+U_{vdw},
\end{equation}
where all the contributions of the pair potentials in Eq. (\ref{eq.U_tot})
are described in detail below.

Excluded volume interactions are modeled via a hard sphere potential
$U_{hs}$ \cite{note_HS} defined as follows
%
\begin{equation}
\label{eq.U_hs}
U_{hs}(r)=\left\{
\begin{array}{ll}
\infty,
& \mathrm{for}~r < r_{cut} \\
0,
& \mathrm{for}~r \geq r_{cut}
\end{array}
\right.
\end{equation}
%
where $r_{cut}=\sigma$ for the microion-microion excluded volume interaction, and
$r_{cut}=a+\sigma/2$ for the macroion-microion excluded volume interaction.
Hence the center-center distance of closest approach between the macroion and
a microion is $r_0=a+\sigma/2=5\sigma$.

The pair electrostatic interaction between two ions $i$ and $j$ (where
$i$ and $j$ can be either a microion or the macroion) reads

\begin{equation}
\label{eq.U_coul} U_{coul}(r_{ij})=\pm k_{B}Tl_{B}\frac{Z_i
Z_j}{r_{ij}},
\end{equation}
%
where +(-) applies to charges of the same (opposite) sign and
$l_{B}=e^{2}/4\pi \epsilon _{0}\epsilon _{r}k_{B}T$ is the Bjerrum
length corresponding to the distance at which two elementary
charges interact with $k_B T$.  To link our simulation parameters
to experimental units and room temperature ($T=298$K) we choose
$\sigma =4.25$ \AA\ leading to the Bjerrum length of water
$l_{B}=1.68\sigma =7.14$ \AA\ and to a macroion surface charge
density of $0.14 \cdot \mathrm{Cm^{-2}}$.

The polyelectrolyte chain connectivity is modeled by using a standard
FENE potential in good solvent (see, e.g., \cite{Kremer_FENE_1993}),
which reads
%
\begin{equation}
\label{eq.U_fene}
U_{FENE}(r)=
\left\{ \begin{array}{ll}
\displaystyle -\frac{1}{2}\kappa R^{2}_{0}\ln \left[ 1-\frac{r^{2}}{R_{0}^{2}}\right] ,
& \textrm{for} \quad r < R_0 \\ \\
\displaystyle \infty ,
& \textrm{for} \quad r \geq R_0 \\
\end{array}
\right.
\end{equation}
%
where we chose $\kappa = 27k_{B}T/\sigma ^{2}$ and
$R_{0}=1.5\sigma $.
The excluded volume interaction between chain monomers is taken into
account via a purely repulsive Lennard-Jones (LJ) potential given
by
%
\begin{equation}
\label{eq.LJ}
U_{LJ}(r)=
\left\{ \begin{array}{ll}
\displaystyle
4\epsilon \left[ \left(\frac{\sigma}{r}\right)^{12}
-\left( \frac{\sigma }{r}\right) ^{6}\right] +\epsilon,
& \textrm{for} \quad r \leq 2^{1/6}\sigma \\ \\
0,
& \textrm{for} \quad  r > 2^{1/6}\sigma
\end{array}
\right.
\end{equation}
%
where $\epsilon=k_BT$.
These parameters lead to an equilibrium bond length $ l=0.98\sigma$.

An important interaction in the multilayering process addressed in
this study is the \textit{non}-electrostatic \textit{short ranged
attraction}, $U_{vdw}$, between the macroion and the PC chain.  To
account for this kind of interaction, we choose without loss of
generality a van der Waals (VDW) potential of interaction between
the macroion and a PC monomer that is given by
\begin{table}[b]
\caption{
Model simulation parameters with some fixed values.
}
\label{tab.simu-param}
\begin{ruledtabular}
\begin{tabular}{lc}
 Parameters&
\\
\hline
 $T=298K$&
 room temperature\\
 $Z_M=40$ &
 macroion valence\\
 $Z=1$&
 microion valence\\
 $\sigma =4.25$ \AA\ &
 microion diameter\\
 $l_{B}=1.68\sigma =7.14$ \AA\ &
 Bjerrum length\\
 $a=4.5\sigma$ &
 macroion radius \\
 $r_0=a+\frac{\sigma }{2}=5\sigma$ &
 macroion-microion distance of closest approach \\
 $R=60\sigma $&
 radius of the outer simulation cell\\
 $N_+$&
 number of PCs\\
 $N_-$&
 number of PAs\\
 $N_{PE}=N_+ + N_-$&
 total number of polyelectrolyte chains\\
 $N_m$&
 number of monomers per chain\\
 $\chi_{vdw}$&
 strength of the specific van der Waals attraction
\end{tabular}
\end{ruledtabular}
\end{table}
%
\begin{equation}
\label{eq.U_vdw}
U_{vdw}(r)=-\epsilon \chi_{vdw}
\left( \frac{\sigma}{r-r_0+\sigma} \right)^6
\hspace{0.5cm} \textrm{for}~ r \geq r_0 ,
\end{equation}
%
where $\chi_{vdw}$ is a positive dimensionless parameter describing the
strength of the attraction. Thereby, at contact (i.e., $r=r_0$),
the magnitude of the attraction is $\chi_{vdw} \epsilon=\chi_{vdw} k_BT$, and
for $\chi_{vdw}=1$, one recovers the standard attractive component of the
LJ-potential [see Eq. \eqref{eq.LJ}].
Since it is not straightforward to link this strength of
adsorption directly to experimental values, we therefore investigated
different possible strengths of attraction, which are known to be realistic for
soft matter systems.

All the simulation parameters are gathered in Table
\ref{tab.simu-param}. The set of simulated systems can be found in
Table \ref{tab.simu-runs}.
Single-particle moves were considered with an acceptance ratio of
$20-30\%$ for the monomers and $50\%$ for the counterions.
At equilibrium, the (average) length of the trial moves $\Delta r$ are about
$30 \sigma$ for the counterions and $0.1 \sigma$ for the monomers.
About $10^5$ to $10^6$ MC steps per
particle were required for equilibration, and about $2 \times
10^6$ subsequent MC steps were used to perform measurements.

\begin{table}[b]
\caption{
System parameters. The number of counterions (cations and anions) ensuring
the overall electroneutrality of the system is not indicated.
}
\label{tab.simu-runs}
\begin{ruledtabular}
\begin{tabular}{lcccc}
 System&
 $N_{PE}$&
 $N_+$&
 $N_-$&
 $N_m$
\\
\hline
 $A$&
 $1$&
 $1$&
 $0$&
 $80$\\
 $B$&
 $2$&
 $1$&
 $1$&
 $80$\\
 $C$&
 $3$&
 $2$&
 $1$&
 $80$\\
 $D$&
 $4$&
 $2$&
 $2$&
 $80$\\
 $E$&
 $5$&
 $3$&
 $2$&
 $80$\\
 $F$&
 $6$&
 $3$&
 $3$&
 $80$\\
 $G$&
 $12$&
 $6$&
 $6$&
 $80$\\
 $H$&
 $40$&
 $20$&
 $20$&
 $10$\\
\end{tabular}
\end{ruledtabular}
\end{table}

\section{Target quantities
 \label{Sec.Target}}

Before presenting the results, we briefly describe the different observables
that are going to be measured.  Of first importance, we compute the radial
density of monomers $n_{\pm}(r)$ around the spherical macroion normalized as
follows

\begin{equation}
\label{eq.n_r}
\int ^{R}_{r_0}4\pi r^2 n_\pm(r)dr=N_\pm N_m
\end{equation}
%
where $(+)-$ applies to PCs (PAs). This quantity is of special interest
to characterize the degree of ordering in the vicinity of the macroion surface.

The total number of  accumulated monomers $\bar{N}_{\pm}(r)$ within a distance $r$
from the macroion center is then given by
%
\begin{equation}
\label{eq.N_r}
\bar{N}_\pm(r) = \int ^{r}_{r_0}4\pi r'^2 n_\pm(r')dr'
\end{equation}
%
where $(+)-$ applies to PCs (PAs). This observable will be used for
the study of the adsorption of (i) a single PC chain
(Sec. \ref{Sec.single-chain}) and (ii) two oppositely charged
polyelectrolytes (Sec. \ref{Sec.two-chains}).

Another quantity of special interest is the global \textit{net fluid
charge} $Q(r)$ which is defined as follows
\begin{equation}
\label{Eq.Qr}
Q(r)=\int ^{r}_{r_0}4\pi r'^{2}Z\left[
\tilde{n}_{+}(r')-\tilde{n}_{-}(r')\right] dr',
\end{equation}
%
where $\tilde{n}_+$ ($\tilde{n}_-$) include the density of all the
positive (negative) microions.  Thus, $Q(r)$ corresponds to the
total fluid charge (omitting the macroion bare charge $Z_M$)
within a distance $r$ from the macroion center, and at the cell
wall $Q(r=R)=Z_M$.  Up to a factor proportional to $1/r^{2}$,
$\left[ Q(r)-Z_M\right]$ gives (by simple application of the Gauss
theorem) the mean electric field at $r$.  Therefore $Q(r)$ can
measure the strength of the macroion charge screening by the
charged species present in its surrounding solution.

\section{Single chain adsorption
 \label{Sec.single-chain}}

In this part, we study the adsorption of a single long PC
chain (system $A$) for different couplings $\chi_{vdw}$.
Experimentally this would correspond to the process of the
\textit{first} polyelectrolyte layer formation.

\begin{figure}
\includegraphics[width = 8.0 cm]{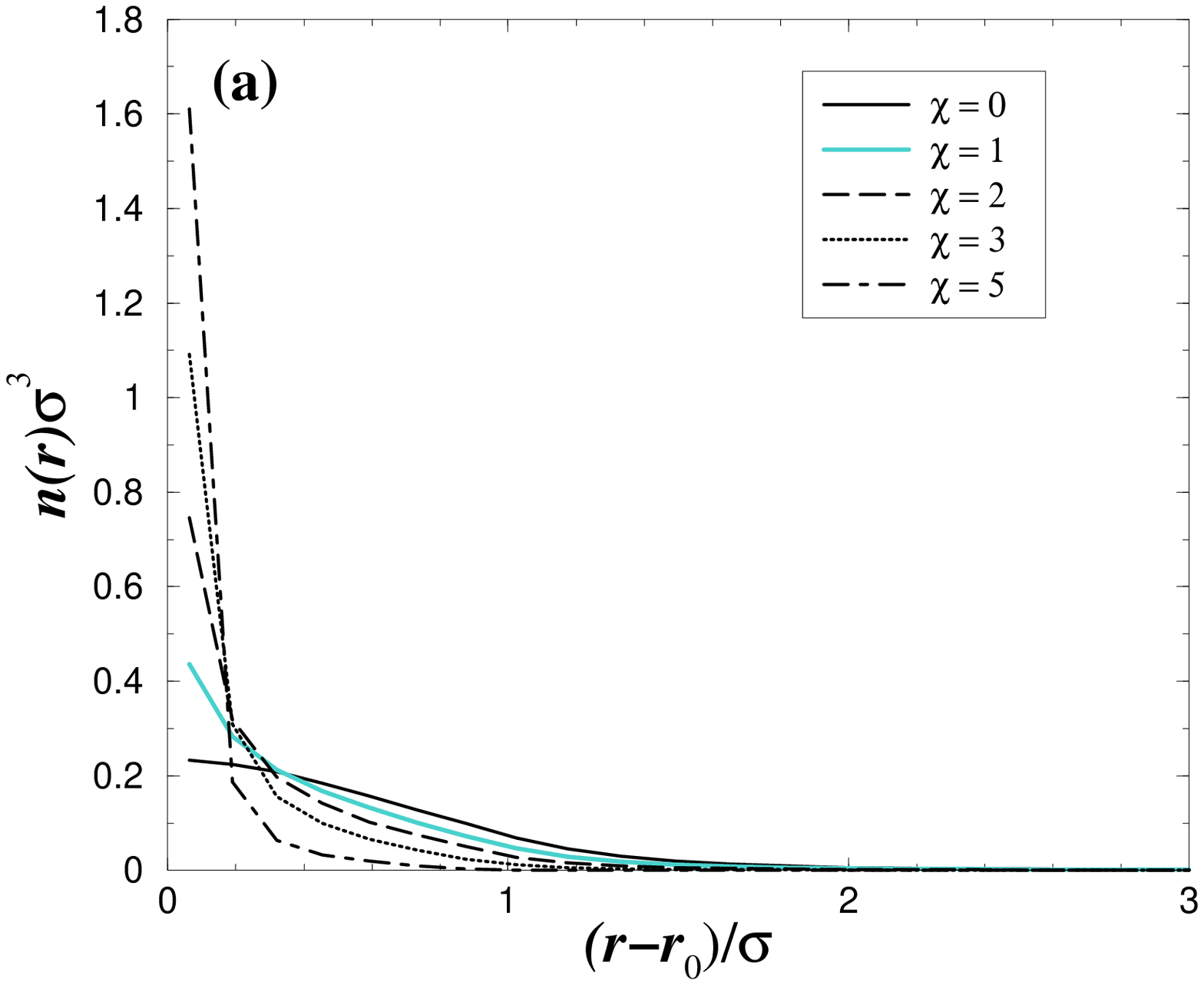}
\includegraphics[width = 8.0 cm]{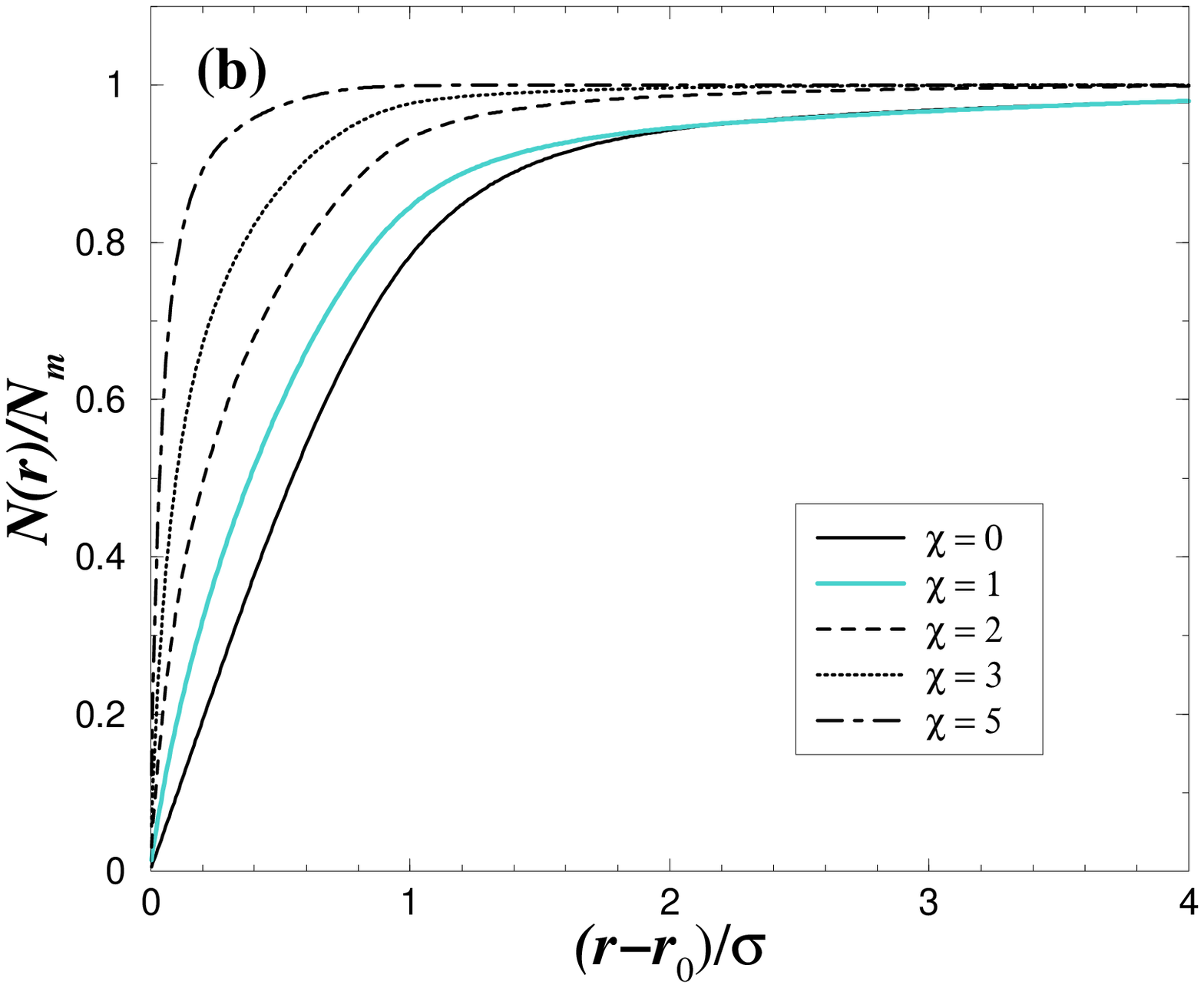}
\caption{
Monomer adsorption profiles of a single PC-chain (system $A$)
at different $\chi_{vdw}$ couplings.
(a) radial density $n_+(r)$.
(b) fraction of monomers $\bar{N}_+(r)/N_m$.
}
\label{fig.nr_SINGLE_CHAIN}
\end{figure}

The monomer density $n_+(r)$ and fraction $\bar{N}_+(r)/N_m$ are
depicted in Fig. \ref{fig.nr_SINGLE_CHAIN}(a) and
Fig. \ref{fig.nr_SINGLE_CHAIN}(b), respectively.  The density $n_+(r)$
near contact ($r \sim r_0$) increases considerably with $\chi_{vdw}$ as
expected. At a radial distance of $1.5\sigma$ from the macroion
surface (i.e., $r=r_0+\sigma$), more than 97\% of the monomers are
adsorbed for sufficiently large $\chi_{vdw}$ (here $\chi_{vdw}>3$) against only
78\% for $\chi_{vdw}=0$.

\begin{figure}[b]
\includegraphics[width = 8.0 cm]{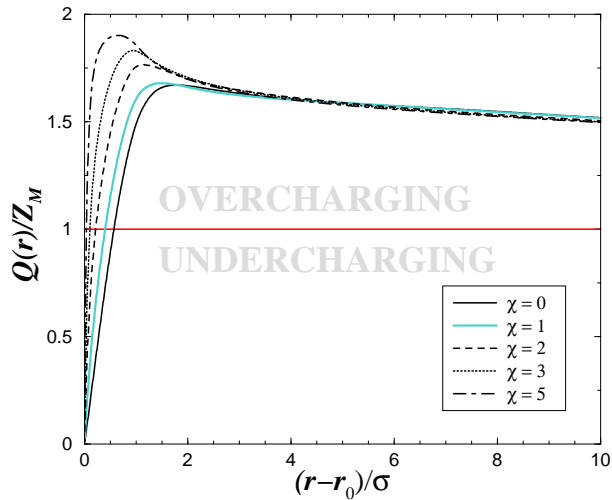}
\caption{ Net fluid charge $Q(r)$ for system $A$ at different $\chi_{vdw}$
couplings.  The horizontal line corresponds to the isoelectric
point.  }
\label{fig.Qr_SINGLE_CHAIN}
\end{figure}

The net fluid charge $Q(r)$ is reported in
Fig. \ref{fig.Qr_SINGLE_CHAIN}.  In all cases we observe a macroion
charge reversal (i.e., $Q(r)/Z_M>1$), as expected from previous
studies \cite{Wallin_Langmuir_1996_I,Wallin_JPhysChem_1996_II}
addressing only $\chi_{vdw}=0$.  The position $r=r^*$ at which $Q(r^*)$ gets
its maximal value decreases with $\chi_{vdw}$, due to the $\chi_{vdw}$-enhanced
adsorption of the chain.  This \textit{overcharging} increases with
$\chi_{vdw}$, since the gain in energy by macroion-monomer VDW interactions
can better overcome (the higher $\chi_{vdw}$) the cost of the self-energy
stemming from the adsorbed excess charge.  Note that the maximal value
of charge reversal of $100\%$ allowed by the total PC charge (i.e.,
$Q(r^*)/Z_M=2$) can not be reached due to a slight accumulation of
microanions.

%
\begin{figure*}
\includegraphics[width = 16.0 cm]{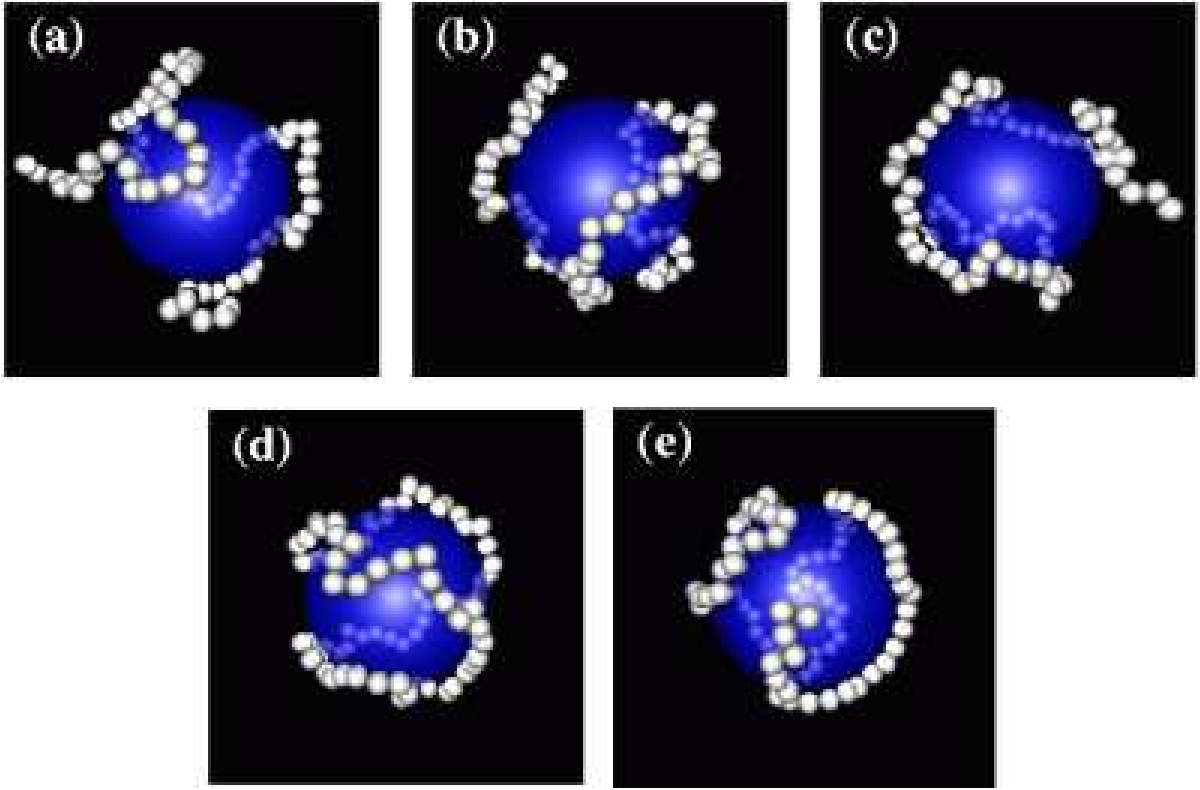}
\caption{
Typical equilibrium configurations for a single PC adsorbed onto an oppositely
charged macroion (system $A$). The little ions are omitted for clarity.
(a) $\chi_{vdw}=0$
(b) $\chi_{vdw}=1$
(c) $\chi_{vdw}=2$
(d) $\chi_{vdw}=3$
(e) $\chi_{vdw}=5$.
}
\label{fig.snap_SINGLE_CHAIN}
\end{figure*}

Typical equilibrium configurations can be found in
Fig. \ref{fig.snap_SINGLE_CHAIN}.  For all $\chi_{vdw}$ values, there is a
wrapping of the chain around the macroion.  In parallel, one can
clearly see that the formation of chain loops is gradually inhibited
by increasing $\chi_{vdw}$.

Although all the obtained results are intuitively easy to understand,
they will turn out to be helpful in order to have a quantitative
analysis of the effect of an extra short-range attraction already on
the level of a single chain adsorption.
%
\section{Adsorption of two oppositely charged polyelectrolytes
 \label{Sec.two-chains}}

We now consider the case where we have additionally a PA chain (system $B$),
so that we have a neutral polyelectrolyte complex (i.e., one PC and one PA).
Experimentally this would correspond to the process of the \textit{second}
polyelectrolyte layer formation (with system $A$ as the initial state).  We
stress the fact that this process is fully reversible for the parameters
investigated in our present study. In particular, we checked that the same
final \textit{equilibrium} configuration is obtained either by (i) starting
from system $A$ and then adding a PA or (ii) starting with no chains and then
adding the two oppositely charged polyelectrolytes, together with their
counterions, simultaneously.

\begin{figure}[t]
\includegraphics[width = 8.0 cm]{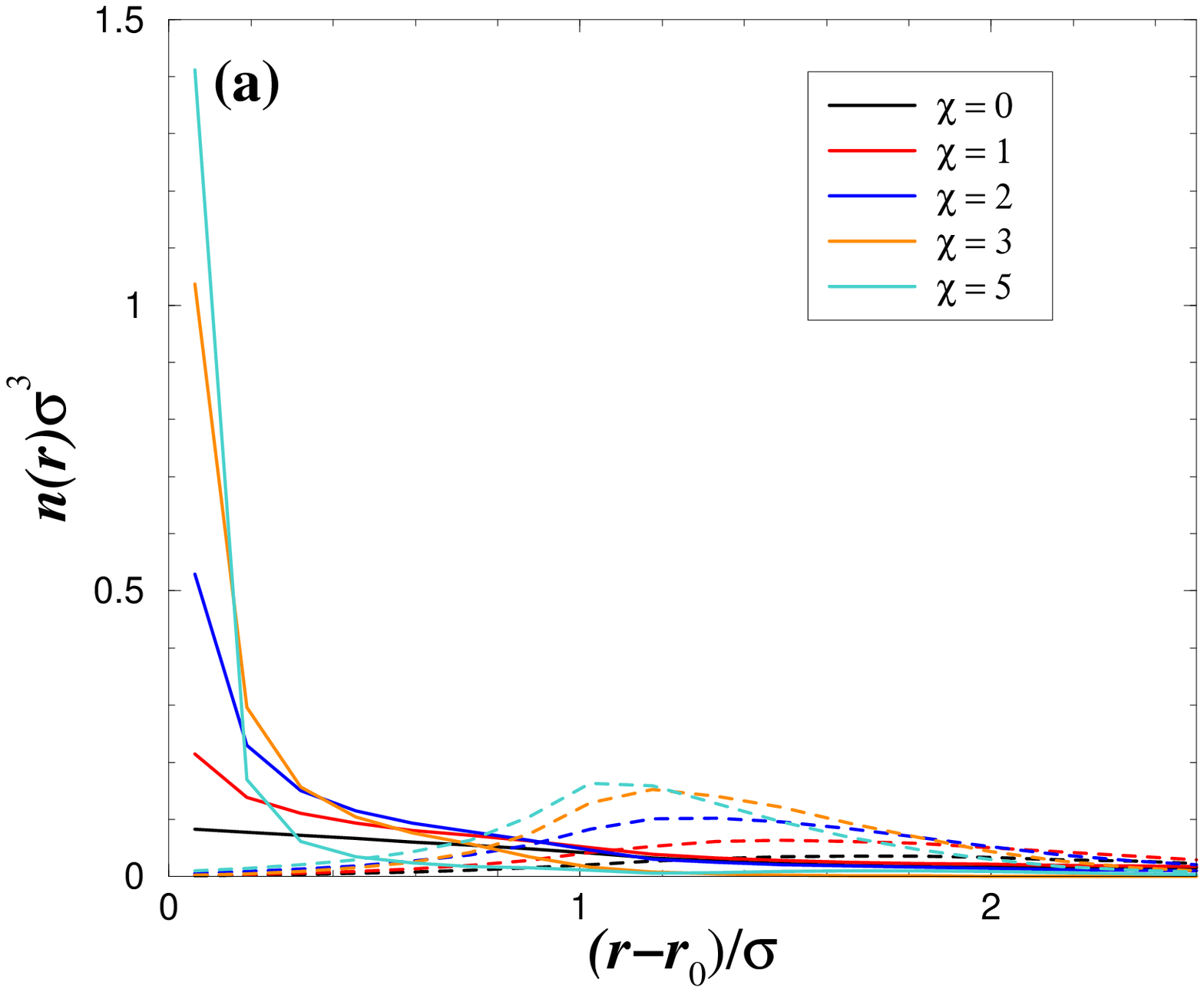}
\includegraphics[width = 8.0 cm]{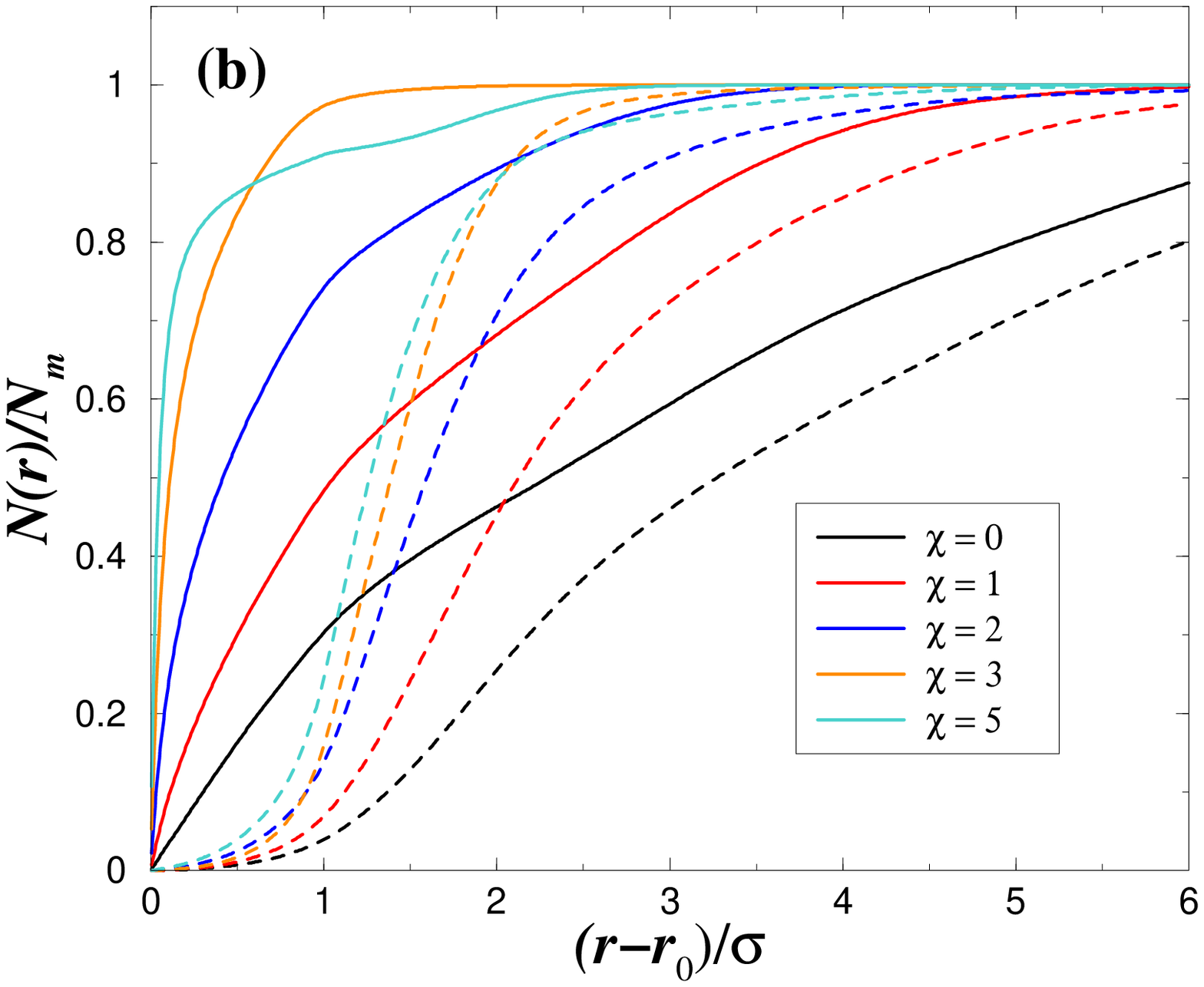}
\caption{
Monomer adsorption profiles of two  polyelectrolyte chains (system $B$)
at different $\chi_{vdw}$ couplings.
The solid and dashed lines correspond to PC and PA monomers, respectively.
(a) radial density $n_\pm(r)$.
(b) fraction of monomers $\bar{N}_\pm(r)/N_m$.
}
\label{fig.nr_TWO_CHAIN}
\end{figure}

\begin{figure*}
\includegraphics[width = 16.0 cm]{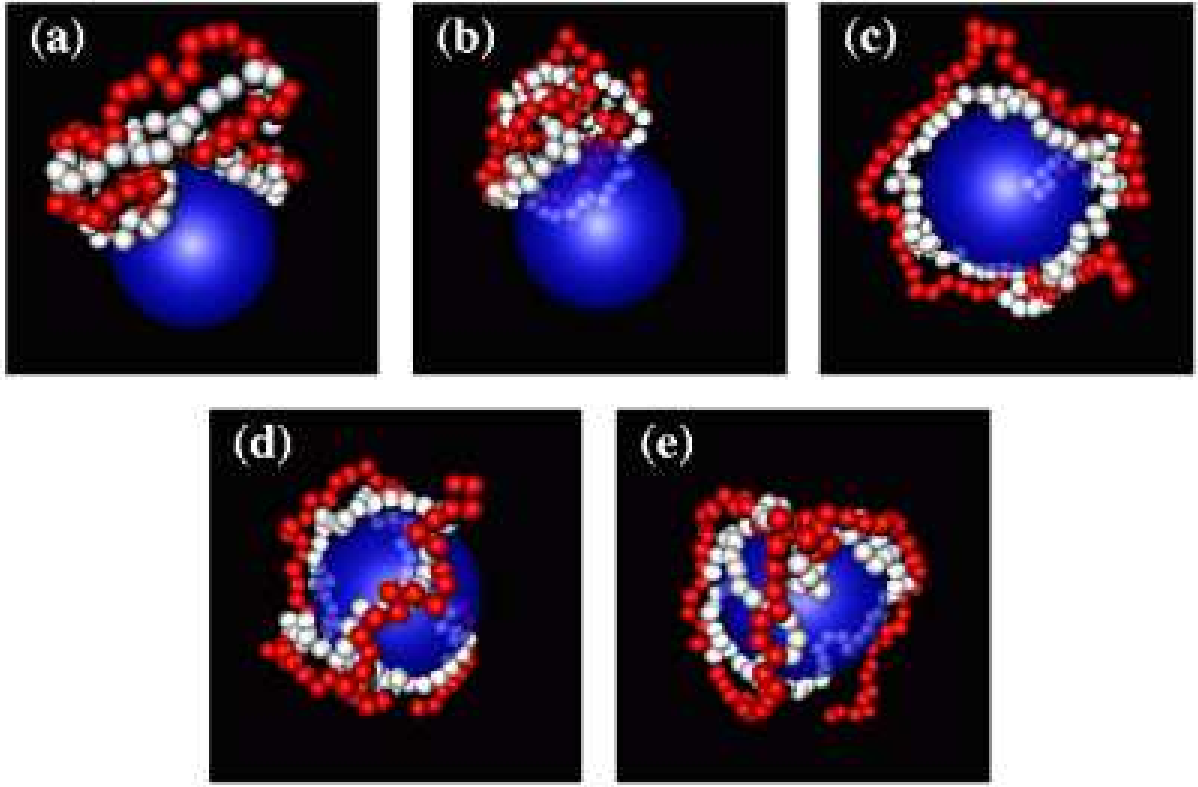}
\caption{
Typical equilibrium configurations for one PC (in white) and one PA
(in red) adsorbed onto the charged macroion (system $B$).
The little ions are omitted for clarity.
(a) $\chi_{vdw}=0$
(b) $\chi_{vdw}=1$
(c) $\chi_{vdw}=2$
(d) $\chi_{vdw}=3$
(e) $\chi_{vdw}=5$.
}
\label{fig.snap_TWO_CHAIN}
\end{figure*}
%

The monomer density $n_\pm(r)$ and fraction $\bar{N}_\pm(r)/N_m$ are
depicted in Fig. \ref{fig.nr_TWO_CHAIN}(a) and
Fig. \ref{fig.nr_TWO_CHAIN}(b), respectively.  The corresponding
microstructures are sketched in Fig. \ref{fig.snap_TWO_CHAIN}.  The
density of PC monomers $n_+(r)$ near contact ($r \sim r_0$) increases
considerably with $\chi_{vdw}$ as expected.  However, a comparison with
system $A$ (see Fig. \ref{fig.nr_SINGLE_CHAIN}) indicates that the
adsorption of PC monomers (at given $\chi_{vdw}$) is weaker when an
additional PA is present. This is consistent with the idea that the PC
chain tends to build up a globular state by getting complexed to the
PA chain. This feature is well illustrated in
Fig. \ref{fig.snap_TWO_CHAIN}.  More precisely, for sufficiently small
$\chi_{vdw} \lesssim 1 $, the polyelectrolyte globular state is highly
favorable compared to the "flat" bilayer state (see also
Fig. \ref{fig.snap_TWO_CHAIN}).  Nevertheless, at sufficiently large
$\chi_{vdw} \gtrsim 2$, the first layer made up of PC monomers is
sufficiently stable to produce a second layer made up of PA
monomers. Thereby, the two chains wrap around the macroion.  As far as
the PA monomer adsorption is concerned, Fig. \ref{fig.nr_TWO_CHAIN}
shows that $n_-(r)$ always increases with $\chi_{vdw}$.  For $\chi_{vdw}=0$, the
polyelectrolyte complex is very close to the globular polyelectrolyte
\textit{bulk} state (i.e., in the absence of the macroion).  This a
non-trivial result, since naively one would expect a "true"
multilayering for any $\chi_{vdw}$. It is only for large $\chi_{vdw} \gtrsim 3$
that one gets a true bilayer formation, where there is a pronounced
peak in $n_-(r)$ around $r-r_0 \approx \sigma$.

It is useful to introduce the following dimensionless interaction
parameters $\Gamma_M = Z_M Z \frac{l_B}{r_0}$, which measures the
strength of the macroion-PC electrostatic attraction, and
$\Gamma_m = Z^2 \frac{l_B}{\sigma}$ which controls the PC-PA
complex interaction. For large values of $\Gamma_m$ the bulk
complex will always be in a globular state, since then the Coulomb
interaction will give rise to a chain collapse, similar to those
seen in polyelectrolyte systems. Thus, for a sufficiently large
value of $\Gamma_m/\Gamma_M$ at given $\chi_{vdw}$, the globular state
will always be favorable and no bilayering can occurs. In this
case \textit{unwrapping} occurs, similarly to the microstructures
depicted in Fig. \ref{fig.snap_TWO_CHAIN}(a) and Fig.
\ref{fig.snap_TWO_CHAIN}(b). On the other hand, we find at fixed
parameters $\Gamma_m$ and $\Gamma_M$, that one needs a s
sufficiently large value $\chi_{vdw}^*$, in order to achieve bilayering.

One can summarize these important results as follows:

\begin{itemize}
\item The equilibrium \textit{bilayering} process on a spherical charged
  colloid with long polyelectrolyte chains requires a sufficiently strong
  extra short-ranged macroion-PC attraction.
\end{itemize}

A closer look on Fig. \ref{fig.nr_TWO_CHAIN}(b) reveals a further
non-trivial behavior in the profiles of $\bar{N}_\pm(r)$ at high
$\chi_{vdw}$.  Very close to the macroion surface we always have a
monotonic behavior of the fraction of adsorbed PC
[$\bar{N}_+(r;\chi_{vdw})$] and PA monomers [$\bar{N}_-(r;\chi_{vdw})$] with
respect to $\chi_{vdw}$ as it should be.  However, for a certain
distance away from the surface we observe an unexpected behavior
where $\bar{N}_+(r;\chi_{vdw}=3)>\bar{N}_+(r;\chi_{vdw}=5)$ as well as
$\bar{N}_-(r;\chi_{vdw}=3)>\bar{N}_-(r;\chi_{vdw}=5)$.  One can qualitatively
explain this effect by the onset of the formation of one (or
several) polyelectrolyte microglobule(s), i.e., small cluster(s)
of oppositely charged monomers [see Figs. \ref{fig.snap_TWO_CHAIN} (d) and (e)].
This is indeed possible because
at high $\chi_{vdw}$ in principle more PC ( and consequently also PA)
monomers want to get close to the macroion surface.
Already for neutral chains a two dimensional flat adsorbed
chain needs a high (surface) binding energy. Compared to bulk conformations
the chain entropy is roughly reduced by $k_BT N \ln(q_{d=2}/q_{d=3})$.
Here $q$ is the effective number of
conformational degrees of freedom of a bond, giving that
$\ln(q_{d=2}/q_{d=3})= O(1)$. Thus local microglobules that induce
a small local desorption, are entropically much more favorable.
Also, on the level of the energy, an increase of $q$ concomitantly
favors the PC-PA microglobule.

\begin{figure}[b]
\includegraphics[width = 8.0 cm]{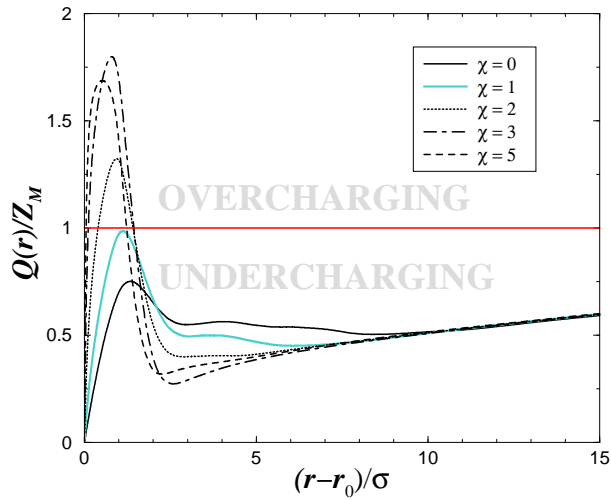}
\caption{
Net fluid charge $Q(r)$ for system $B$ at different $\chi_{vdw}$ couplings.
The horizontal line corresponds to the isoelectric point.
}
\label{fig.Qr_TWO_CHAIN}
\end{figure}

The net fluid charge $Q(r)$ is reported in Fig.\ref{fig.Qr_TWO_CHAIN}.
For $\chi_{vdw}\gtrsim 2$ the macroion gets even overcharged and undercharged as
one gets away from its surface, whereas for low $\chi_{vdw}$ no
\textit{local} overcharging occurs.
Again, at high $\chi_{vdw}$ the strength of the charge oscillation is not a monotonic function
of $\chi_{vdw}$ where we observe a higher local overcharging (and
undercharging) with $\chi_{vdw}=3$ than with $\chi_{vdw}=5$. This latter
feature is fully consistent with the profiles of $\bar{N}_\pm(r)$
[see Fig. \ref{fig.nr_TWO_CHAIN}(b)] previously discussed
\cite{note_counterions_OC}.  However this onset of local (surface)
microglobules (for $\chi_{vdw}=5$) is not strong enough to produce a
non-monotonic behavior of $r^*$ with respect to $\chi_{vdw}$.

\section{Multilayering
 \label{Sec.Multilayer}}

We now turn to the case where there are many polyelectrolytes (with
$N_{PE} \geq 3$) in the system. We recall that when $\chi_{vdw} \neq 0$,
the VDW interaction concern monomers of all PCs lying within the range of interaction.
To keep the number of plots manageable,
we will present our results obtained for $\chi_{vdw}=0$ and $\chi_{vdw}=3$.  The
case $\chi_{vdw}=0$ is (conceptually) important since it corresponds to the
situation where only purely electrostatic interactions are present.
The other case $\chi_{vdw}=3$ seems to be a reasonable choice, since we
found a "true" bilayering for that value.  Moreover, such a strength
should be easily accessible experimentally.

\subsection{Adsorption with $\chi_{vdw}=0$}

\begin{figure}[b]
\includegraphics[width = 8.0 cm]{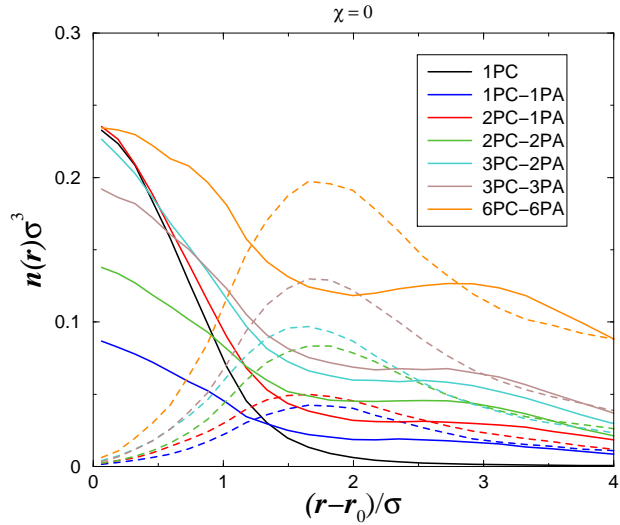}
\caption{
Radial monomer density for the systems $A-G$ with $\chi_{vdw}=0$.
The solid and dashed lines correspond to $n_+(r)$ and $n_-(r)$, respectively.
The number of PC and PA chains is indicated.
The plots of $n_\pm(r)$ for the systems $A$ (1PC) and $B$ (1PC-1PA)
are again reported here for direct comparison.
}
\label{fig.nr_N_CHAIN_VDW0}
\end{figure}

\begin{figure*}
\includegraphics[width = 16.0 cm]{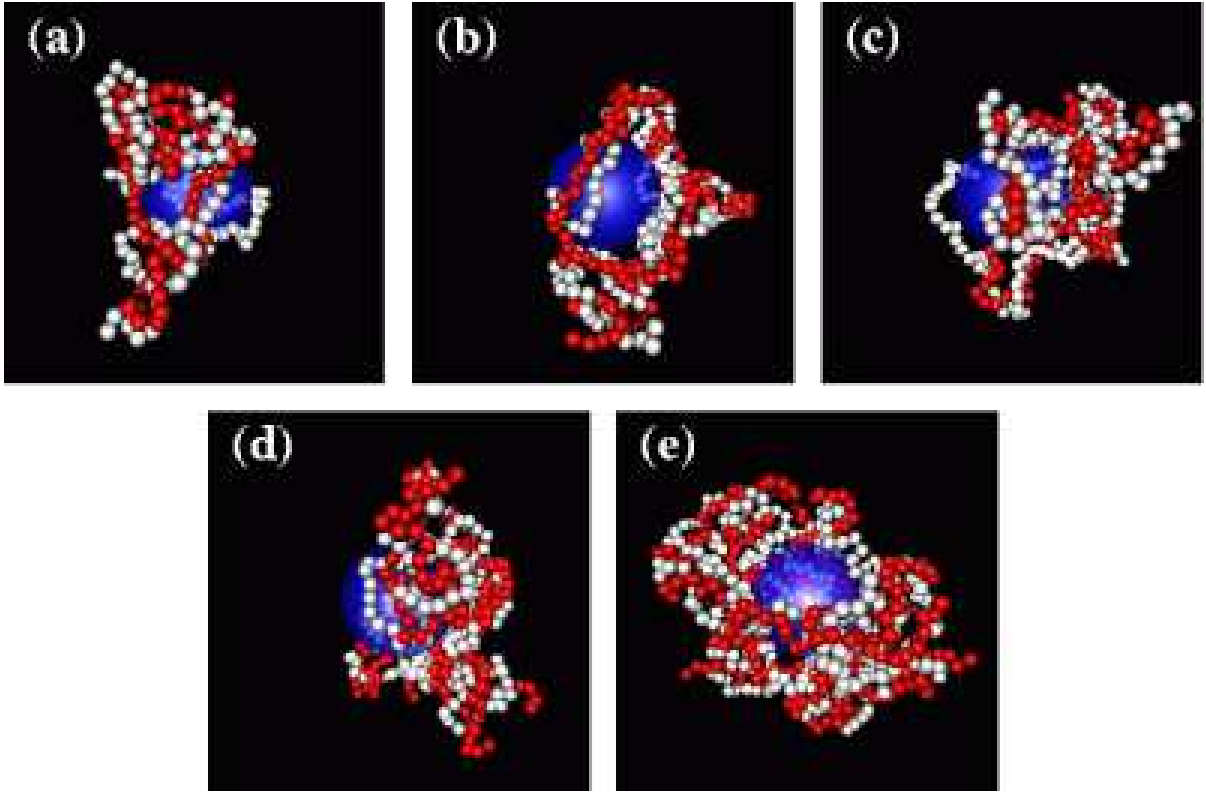}
\caption{
Typical equilibrium configurations for \textit{many} polyelectrolyte chains
adsorbed onto the charged macroion with $\chi_{vdw}=0$.
The PC monomers are in white and  PA ones in red.
The little ions are omitted for clarity.
(a) 2PC-1PA (system $C$)
(b) 2PC-2PA (system $D$)
(c) 3PC-2PA (system $E$)
(d) 3PC-3PA (system $F$)
(e) 6PC-6PA (system $G$).
}
\label{fig.snap_N_CHAIN_VDW0}
\end{figure*}

The density profiles of $n_\pm(r)$  for the systems $A-G$ (with $\chi_{vdw}=0$) are
reported in Fig. \ref{fig.nr_N_CHAIN_VDW0} and the corresponding microstructures
are sketched in Fig. \ref{fig.snap_N_CHAIN_VDW0}.

%

Figure \ref{fig.nr_N_CHAIN_VDW0} shows that when the total polyelectrolyte charge,
%
\begin{equation}
\label{Eq.Q_PE}
Q_{PE} \equiv (N_+-N_-)N_me,
\end{equation}
%
is zero, the density of PC monomers $n_+(r)$ near contact is lower
than when charge $Q_{PE}=N_me$ (recalling that our systems are such
that $Q_{PE}=0$ or $N_me$).  This tendency [lower $n_+(r)$ near
contact with $Q_{PE}=0$] gradually decreases as the total number
$N_{PE}$ of polyelectrolytes is increased.  In particular for the
system $G$ where $N_{PE}=12$ and $Q_{PE}=0$, the density $n_+(r)$ near
contact is nearly identical to that of systems $A$, $C$ and $E$ where
$Q_{PE}=N_me$.  On the other hand, when $Q_{PE} = N_me$ then $n_+(r)$
near contact is nearly independent of $N_{PE}$. The height of the peak
in the PA monomer density $n_-(r)$ \textit{increases monotonically}
with $N_{PE}$.  Concomitantly, a \textit{third} layer made of PC
monomers builds up for high enough $N_{PE}$. This multilayering is
especially remarkable for $N_{PE}=12$ (system $G$).

All these features concerning the \textit{first} layer structure can
be rationalized with simple ideas.  When $Q_{PE}=0$, then the
resulting global attraction between the macroion and the
polyelectrolyte complex is much weaker than when $Q_{PE}=N_me$.  In
this latter situation where $Q_{PE}=N_me$, this excess charge carried
by a PC chain leads to a relatively strong PC adsorption near the
surface.  The equilibrium configurations sketched in
Fig. \ref{fig.snap_N_CHAIN_VDW0} suggest a wrapping from the PC
monomers when $Q_{PE}=N_me$ [see Fig. \ref{fig.snap_N_CHAIN_VDW0}(a)
and (c)] and a (partial) \textit{unwrapping} when $Q_{PE}=0$ [see
Fig. \ref{fig.snap_N_CHAIN_VDW0}(b), (d)].  Even for high $N_{PE}=12$
[see Fig. \ref{fig.snap_N_CHAIN_VDW0}(e)], we can see this tendency of
unwrapping leading to a polyelectrolyte-complex globular state.
However, for symmetry reasons, when the total number of monomers is
large enough as it is the case with $N_{PE}=12$, the distribution of
the polyelectrolyte complex gets more isotropic leading to a weaker
unwrapping at $Q_{PE}=0$.

The collapse into a globular polyelectrolyte complex becomes even more
spectacular when $\sigma$ is reduced (i.e., larger $\Gamma_m$)
\cite{MD_layering}. In that case (not reported here), we found a
wrapping (for $Q_{PE}=N_me$) similar to that depicted in
Fig. \ref{fig.snap_N_CHAIN_VDW0}(a) and (c), and a \textit{strong}
unwrapping (for $Q_{PE}=0$) where the compact neutral polyelectrolyte
complex is adsorbed onto a small area of the colloid.

\begin{figure}[b]
\includegraphics[width = 8.0 cm]{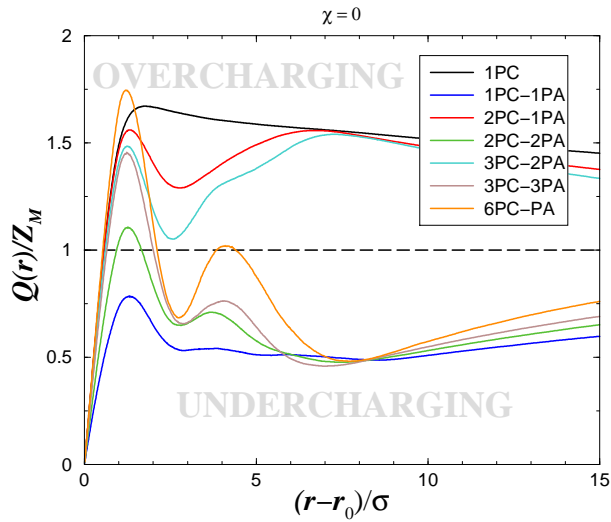}
\caption{
Net fluid charge $Q(r)$ for the systems $A-G$ with $\chi_{vdw}=0$.
The number of PC and PA chains is indicated.
The plots for the systems $A$ (1PC) and $B$ (1PC-1PA)
are again reported here for direct comparison.
The horizontal line corresponds to the isoelectric point.
}
\label{fig.Qr_N_CHAIN_VDW0}
\end{figure}
%

The net fluid charge $Q(r)$ is reported in
Fig. \ref{fig.Qr_N_CHAIN_VDW0}.  As expected one detects an
overcharging and undercharging for $Q_{PE}=N_me$ and $Q_{PE}=0$,
respectively.  For $Q_{PE}=0$, the macroion is also locally
overcharged very close to the macroion surface and its strength
increases with $N_{PE}$.  On the other hand, the strength of the
undercharging (occurring at the largest radial position $r^*$ of the
extrema) at $Q_{PE}=0$ is nearly independent of $N_{PE}$.  In
parallel, the strength of the overcharging (occurring at the largest
radial position $r^*$ of the extrema) measured at $Q_{PE}=N_me$ is
also nearly independent of $N_{PE}$ (systems $C$ and $E$).  Moreover,
our simulations show that for $N_{PE} \geq 2$ the strength of the
overcharging (with $Q_{PE}=N_me$) and undercharging (with $Q_{PE}=0$)
have nearly the same amplitude, in qualitative agreement with
experimental data.

\subsection{Adsorption with $\chi_{vdw} \neq 0$}

In this part, we consider the additional attractive VDW macroion-PC
monomer interaction with $\chi_{vdw}=3$. The same investigation as with
$\chi_{vdw}=0$ is carried out here.

\begin{figure}[b]
\includegraphics[width = 8.0 cm]{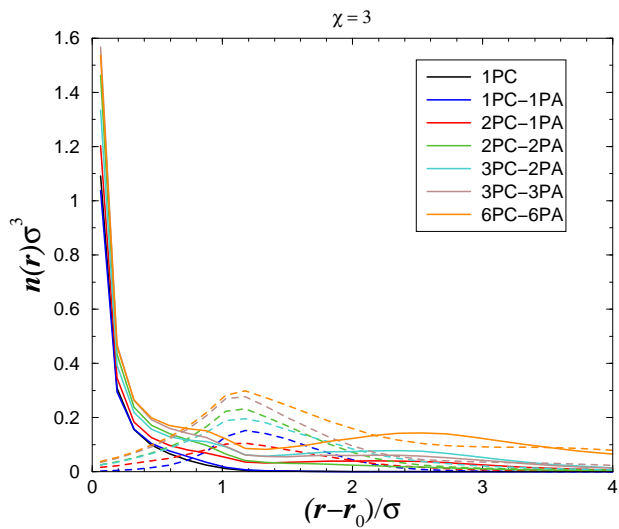}
\caption{
Same as Fig. \ref{fig.nr_N_CHAIN_VDW0} but with $\chi_{vdw}=3$.
}
\label{fig.nr_N_CHAIN_VDW3}
\end{figure}

\begin{figure*}
\includegraphics[width = 16.0 cm]{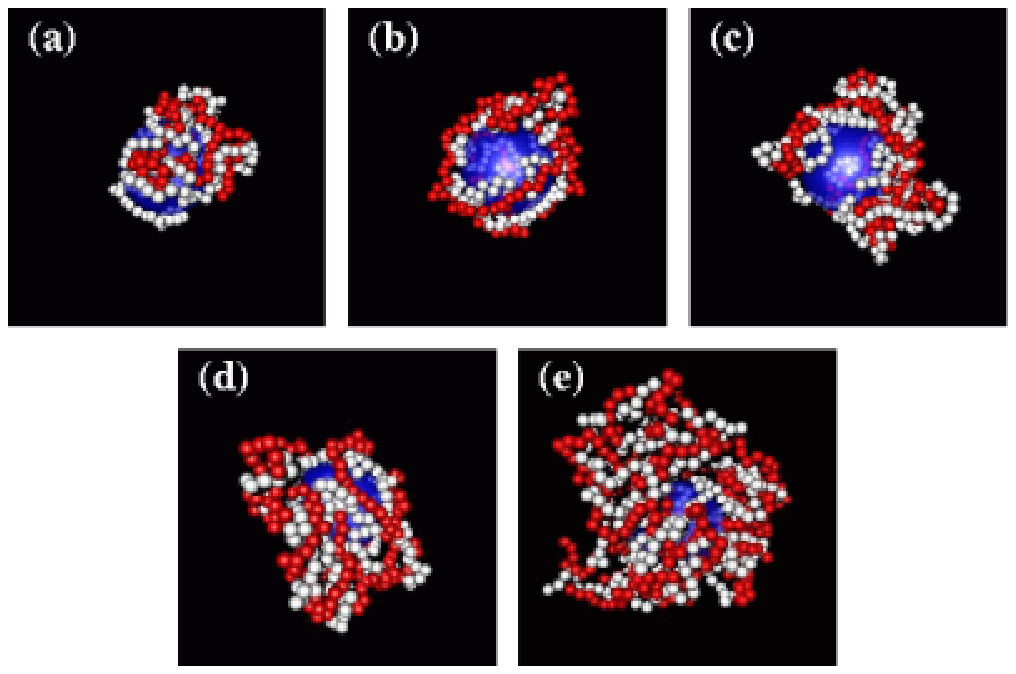}
\caption{
Same as Fig. \ref{fig.snap_N_CHAIN_VDW0} but with $\chi_{vdw}=3$.
(a) 2PC-1PA (system $C$)
(b) 2PC-2PA (system $D$)
(c) 3PC-2PA (system $E$)
(d) 3PC-3PA (system $F$)
(e) 6PC-6PA (system $G$).
}
\label{fig.snap_N_CHAIN_VDW3}
\end{figure*}

The density profiles of $n_\pm(r)$ for the systems $A-G$ (with
$\chi_{vdw}=3$) are reported in Fig. \ref{fig.nr_N_CHAIN_VDW3} and the
corresponding microstructures are sketched in
Fig. \ref{fig.snap_N_CHAIN_VDW3}.
%
%
%
%

Figure \ref{fig.nr_N_CHAIN_VDW3} shows that the density $n_+(r)$ near
contact (for a given system) is about six times larger than that
obtained at $\chi_{vdw}=0$ (compare Fig. \ref{fig.nr_N_CHAIN_VDW0}).  When
$Q_{PE}=N_me$, the density $n_+(r)$ at contact (slightly) increases
monotonically with $N_{PE}$ in contrast to what happened at $\chi_{vdw}=0$
where it was nearly independent of $N_{PE}$.
When $Q_{PE}=0$, we remark that the density $n_+(r)$ near contact is
nearly independent of $N_{PE}$ (for $N_{PE} \geq 2$) in contrast to
what happened at $\chi_{vdw}=0$.

As far as the PA density $n_-(r)$ is concerned, the height of the
first peak (for a given system) is about twice larger than that
obtained at $\chi_{vdw}=0$.  This height is a monotonic function of $N_{PE}$
\textit{within} a given regime of $Q_{PE}$ (here, either $0$ or
$N_me$).  Nevertheless, in general this height exhibits a non-trivial
dependence on $N_{PE}$, in contrast to our results with $\chi_{vdw}=0$.  For
the systems $B$ and $C$ both containing a single PA chain ($N_-=1$),
the height of the first peak in $n_-(r)$ is smaller with $N_+=2$
(system $C$) than with $N_+=1$ (system $B$).  This is again due to the
formation of clusters of oppositely charged monomers that takes place
\textit{above} the first layer. This effect is more pronounced when
the amount of PC monomers (at given $N_-$) is larger (system $C$),
leading to a local desorption of PA monomer.  Those features are
remarkable by a comparison of the snapshots of the systems $B$ and $C$
depicted in Fig. \ref{fig.snap_TWO_CHAIN}(d) and
Fig. \ref{fig.snap_N_CHAIN_VDW3}(a), respectively.  Similar arguments
can be used for the systems $D$ and $E$, where the same effect is
found.  At large $N_{PE}$, the height of the first peak in $n_-(r)$
saturates as expected.

For $3 \leq N_{PE} \leq 6$, our simulation shows that the formation of
the \textit{third} layer [i.e., the second peak in $n_+(r)$ at $r-r_0
\approx 2.6 \sigma$] is enhanced when $Q_{PE}=N_me$ at fixed $N_+$.
This effect can again be explained in terms of polyelectrolyte
(micro)globules formation.  Indeed, above the \textit{second} layer,
the formation of clusters made up of oppositely charged monomers is
enhanced when the polyelectrolyte complex (seen by the underneath
bilayer) is uncharged which corresponds to a state of charge
$Q_{PE}=0$.

It is interesting to see that with $N_{PE}=12$ one even gets a second
peak (and not a shoulder) in $n_-(r)$, which is the signature of a
\textit{fourth} layer.  This qualitatively contrasts with our findings
at $\chi_{vdw}=0$.  Therefore, we conclude that the effect of an extra
short-ranged macroion-PC attraction is crucial for the multilayering
process.

On a more qualitative level, it is very insightful to compare the
microstructures obtained with purely electrostatic interactions
($\chi_{vdw}=0$) sketched in Fig. \ref{fig.snap_N_CHAIN_VDW0} with those
obtained with a short-ranged VDW macroion-PC interaction ($\chi_{vdw}=3$)
sketched in Fig. \ref{fig.snap_N_CHAIN_VDW3}. From such a visual
inspection, it is clear that in all cases the adsorbed polyelectrolyte
complex is flatter at $\chi_{vdw}=3$ than at $\chi_{vdw}=0$.  An other important
qualitative difference, is that the \textit{unwrapping} occurring at
$\chi_{vdw}=0$ with $Q_{PE}=0$ [see Fig. \ref{fig.snap_N_CHAIN_VDW0}(b) and
(d)] is no longer effective when $\chi_{vdw}=3$ [see
Fig. \ref{fig.snap_N_CHAIN_VDW3}(b) and (d)].
In the same spirit, for
a large number of chains ($N_{PE}=12$), the macroion surface is only
partially covered by the PC monomers where some (large) holes appear
[see Fig. \ref{fig.snap_N_CHAIN_VDW0}(e)], in contrast to what occurs
at $\chi_{vdw}=3$, where all the macroion surface is covered [see
Fig. \ref{fig.snap_N_CHAIN_VDW3}(e)].
\begin{figure}[t]
\includegraphics[width = 8.0 cm]{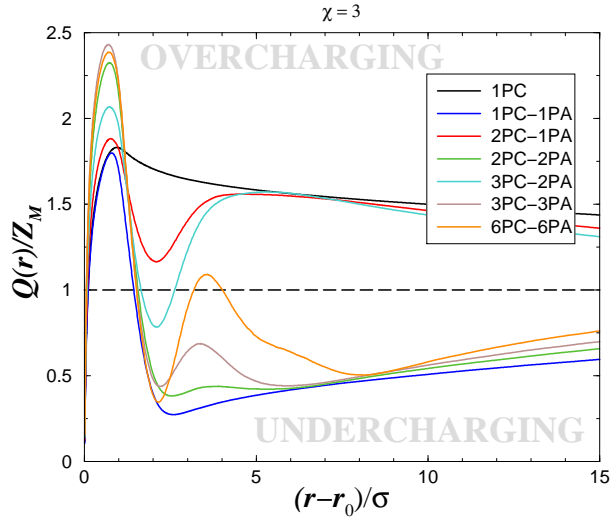}
\caption{
Same as Fig. \ref{fig.Qr_N_CHAIN_VDW0} but with $\chi_{vdw}=3$.
}
\label{fig.Qr_N_CHAIN_VDW3}
\end{figure}

The net fluid charge $Q(r)$ is reported in Fig. \ref{fig.Qr_N_CHAIN_VDW3}.  As
expected one finds an overcharging and undercharging for $Q_{PE}=N_me$ and
$Q_{PE}=0$, respectively.  Now one can get a local overcharging larger than
100\% (i.e., $Q(r)/Z_M>2$) due to the VDW attraction that can lead to a first
layer with \textit{many} PC chains.  For systems $C$ and $E$ we see that the
overcharging at the third layer is around 50\% and nearly independent of
$N_{PE}$.

On the other hand, the strength of the undercharging (occurring at
the largest radial position $r^*$ of the extrema) at $Q_{PE}=0$
decreases with increasing $N_{PE}$, providing a gradually weaker
driving force for the subsequent adsorption of the PC chain.  On
the basis of our results with $\chi_{vdw}=0$, we expect that the
strength of the undercharging at $\chi_{vdw}=3$ (for larger $N_{PE}$)
should stabilize around 50\%.  So it appears that the oscillation
of under- and overcharging are not 100\%, but rather close to 50\%.
This is probably sensitive to the specific model parameters chosen.
What can be stated from our data is, that there is no reason to
find a generally applicable overcharging fraction. Especially for
the case of relatively small colloids, the results will strongly
depend on the specific system parameters, which are both of
electrostatic as well as non-electrostatic nature.

\section{Case of short chains
 \label{Sec.short-chains}}

\begin{figure}
\includegraphics[width = 8.0 cm]{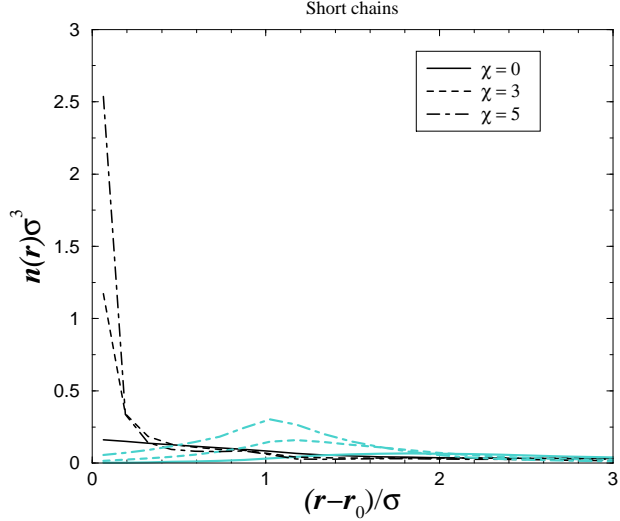}
\caption{
Radial monomer density for short polyelectrolyte chains (system $H$)
at different $\chi_{vdw}$ couplings.
The black and gray lines correspond to PC and PA monomers, respectively.
}
\label{fig.nr_N_SHORT_CHAIN}
\end{figure}

\begin{figure*}[b]
\includegraphics[width = 17.0 cm]{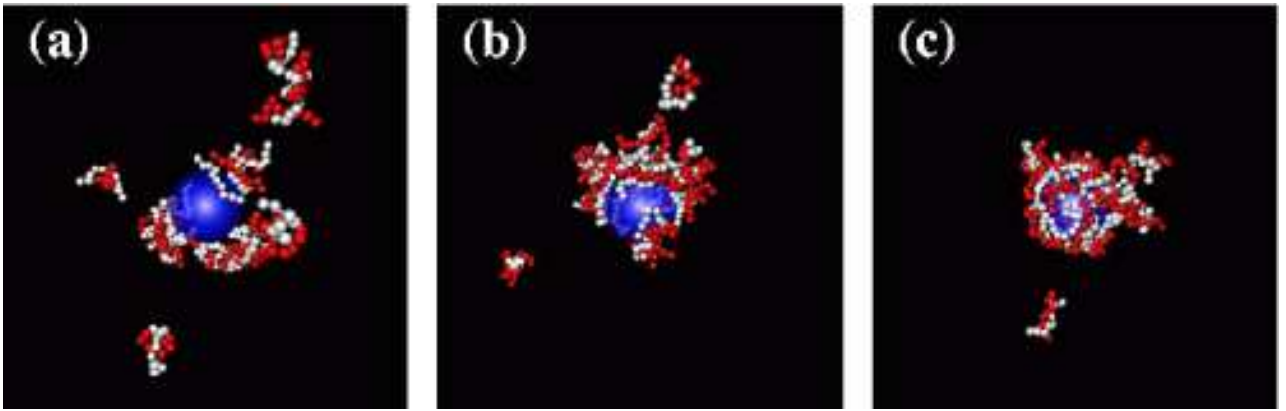}
\caption{
Typical equilibrium configurations for short PC (in white) and PA
(in red) chains adsorbed onto the charged macroion (system $H$).
The little ions are omitted for clarity.
(a) $\chi_{vdw}=0$
(b) $\chi_{vdw}=3$
(c) $\chi_{vdw}=5$
}
\label{fig.snap_N_SHORT_CHAIN}
\end{figure*}

We now investigate the effect of chain length dependence. In this case
the adsorption of a single chain does not necessarily produce an
overcharging since the chain length ($N_m=10$ - system $H$) is too
short.  The density profiles of $n_\pm(r)$ are reported in
Fig. \ref{fig.nr_N_SHORT_CHAIN} for various $\chi_{vdw}$, and the
corresponding microstructures are sketched in
Fig. \ref{fig.snap_N_SHORT_CHAIN}.
%
%
%
In the purely electrostatic regime ($\chi_{vdw}=0$), the polyelectrolyte
adsorption is weak and it significantly increases with $\chi_{vdw}$.
However, for all reported cases, we only observe a bilayering in
contrast with previous long chain systems (compare
Fig. \ref{fig.nr_N_SHORT_CHAIN} with Fig. \ref{fig.nr_N_CHAIN_VDW0}
and Fig. \ref{fig.nr_N_CHAIN_VDW3}) where thereby a true multilayering
was reported.

In addition we observe several globally neutral polyelectrolyte complexes in
the bulk, whose number decreases with $\chi_{vdw}$ (see Fig.
\ref{fig.snap_N_SHORT_CHAIN}). This feature was inhibited for long chains due
to the strong PC-PA binding energy that keeps all the chains near the macroion
surface.  At sufficiently strong $\chi_{vdw}$ [see Fig.
\ref{fig.snap_N_SHORT_CHAIN}(c) with $\chi_{vdw}=5$], the macroion area gets
largely (and \textit{uniformly}) covered by the PC chains, leading to a strong
bilayering. Nevertheless, due to the weak PC-PA binding energy, the formation
of additional layer seems to be prohibited in contrasts to what was observed
at $\chi_{vdw}=3$ with systems $D$ and $E$ that contain a similar number of
monomers.  Those observations lead us to the relevant conclusion that
multilayering with short chains (if experimentally observed on a charged
colloidal sphere) must involve additional non-trivial driving forces like
specific PC monomer-PA monomer interactions that are not captured by our
model. This again seems to be in agreement with the arguments presented in
Ref.\cite{Kovacevic_2002} which argue against a stable thermodynamic
equilibrium complex when there is excess polyelectrolytes present.

\begin{figure}
\includegraphics[width = 8.0 cm]{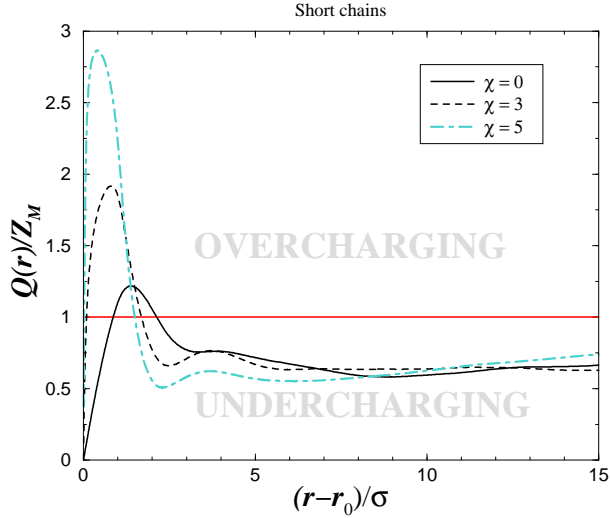}
\caption{
Net fluid charge $Q(r)$ for system $H$ at different $\chi_{vdw}$ couplings.
The horizontal line corresponds to the isoelectric point.
}
\label{fig.Qr_N_SHORT_CHAIN}
\end{figure}

The net fluid charge plotted in Fig. \ref{fig.Qr_N_SHORT_CHAIN}
indicates that only one charge oscillation (around the isoelectric
point) is obtained in contrast to what can happen with longer
chains.  Again, here the driving force for the bilayer formation is
the overcharging that increases with $\chi_{vdw}$.

\section{Concluding remarks
 \label{Sec.conclu}}

We have carried out MC simulations to study the basic mechanisms involved in
forming equilibrium polyelectrolyte complexes on a charged colloidal sphere.
This work emphasizes the role of the short-range Van der Waals-like attraction
(characterized here by $\chi_{vdw}$) between the spherical macroion surface
and the oppositely charged adsorbed chain(s).

It was demonstrated that, for the bilayering process involving two long
oppositely charged chains at fixed $\Gamma_m$ and $\Gamma_M$, it is necessary
to have a sufficiently high $\chi_{vdw}$.  In particular, below a certain
value of $\chi_{vdw}=\chi_{vdw}^*$, a dense adsorbed polyelectrolyte globule
is obtained, whereas above $\chi_{vdw}^*$ a flat bilayer builds up.

The same qualitatively applies to the case of many (more than two) long
polyelectrolytes.  In a purely electrostatic regime (i.e., $\chi_{vdw}=0$) one
can never obtain a true (uniform) multilayering.  However, by increasing
$\chi_{vdw}$, one gradually increases the polyelectrolyte (polycation and
polyanion) chain adsorption ultimately leading to a true multilayering where
the macroion is uniformly covered.  Nonetheless, at given $\chi_{vdw}$ and
especially for small $\chi_{vdw}$, the polyelectrolyte globular state is
always favored when its net charge is zero.

As far as the short chain case is concerned, it was shown that even bilayering
can not be reached within the pure electrostatic regime. Only at higher
$\chi_{vdw}$ (higher than those coming into play with long chains), one
recovers a bilayer formation.  However, multilayering (beyond bilayering) with
very short chains seems to be very unlikely within our model. The large
complex would not be thermodynamically stable and dissolve into smaller charge
neutral polyelectrolyte complexes, consistent with the ideas presented in
Ref.\cite{Kovacevic_2002}.

As an overall conclusion, our results clearly demonstrated that besides an
overcharging driving force [i.e., successive macroion (effective) charge
reversal by successive polymer layering), the stability of the polyelectrolyte
multilayer is strongly influenced by the specific macroion-polyelectrolyte
short range attraction.  This statement should at least hold for the
investigated cases of equilibrium structures.

A future study should systematically study other important effects, such as
chain flexibility, specific interchain monomer-monomer interaction, microions
valency etc...  Nevertheless, our present findings hopefully will generate
some further systematic studies to explore the effects of non-electrostatic
effects for the layer-by-layer deposition technique.

\acknowledgments R. M. thanks F. Caruso and S. K. Mayya for helpful and
stimulating discussions.  This work was supported by \textit{Laboratoires
  Europ\'eens Associ\'es} (LEA) and the SFB 625.


\begin{thebibliography}{10}
\bibitem{note_HS}
Only the monomer-monomer excluded volume interaction was not modelled by a
  hard-sphere potential. There, a purely repulsive Lennard-Jones potential was
  used.
\bibitem{note_counterions_OC} Here we assume the presence of little
  counterions to be negligible in the vicinity of the macroion surface, which
  is a very good approximation.
\end{thebibliography}
\providecommand{\refin}[1]{\\ \textbf{Referenced in:} #1}

\end{document}